# Using a Smartphone pressure sensor as Pitot tube speedometer

D. Dorsel, S. Staacks, H. Heinke, C. Stampfer
*Institute of Physics I and II, RWTH Aachen University*

## Introduction

As smartphones have become a part of our everyday life, their sensors have successfully been used to allow data acquisition with these readily available devices in a variety of different smartphone-based school experiments[1-4]. Such experiments most commonly take advantage of the accelerometer and gyroscope. A less frequently used sensor in smartphone-based experiments is the pressure sensor or barometer[5-7]. Pressure sensors in smartphones can improve the indoor navigation, for example in multi-story shopping malls. In a popular smartphone experiment, the barometer is used to determine the current altitude in an elevator with the barometric height formula[8]. Along with accelerometer data and by deriving the height data to calculate the velocity, z(t), v(t) and a(t) plots can be generated and shown to students in real-time[9].

In this article, an experiment to measure air velocity with the barometer of a smartphone is presented utilizing the concept of a Pitot tube. For the experiment, the app phyphox is used which has a set of useful features for performing smartphone-based experiments[9,10]. Both the data analysis tools included in the app phyphox and the remote access function to smartphones in experimental setups are used in the presented experiment. This experiment is also an example for designing customized experiments within phyphox which can be realized by any user via a web-based editor[9]. In addition, it is shown how an external Bluetooth pressure sensor can be used to extend an experiment, which is used here to measure the speed of a vehicle. Such an integration of external Bluetooth Low Energy sensors into smartphone experiments became available with phyphox version 1.1.0.

## The Pitot tube

Fluid flow velocities can be measured with a Pitot tube[11] (Figure 1). One opening of the Pitot tube points against the flow direction where it is subject to stagnation pressure. The other opening is perpendicular to the flow direction, which in contrast makes the static pressure accessible. Original Pitot tubes contain water, which ascends to different heights due to the pressure difference.

The pressure difference depends on the fluid velocity and can be expressed by the Bernoulli equation:

$$p_{static} + \rho_{fluid} \cdot \frac{v^2}{2} = p_{stagnation} \qquad (1)$$

Rearranging the terms allows calculating the velocity from the pressure difference.

$$v = \sqrt{\frac{2 \cdot (p_{stagnation} - p_{static})}{\rho_{fluid}}} \qquad (2)$$

In general, the Bernoulli equation is valid for incompressible fluids. Air can be considered as incompressible for velocities smaller than 30% of the speed of sound[12]. The temperature dependence of the fluid density $\rho_{fluid}$ can be taken into account by measuring the air temperature $T$ and using the known function $\rho_{Air}(T)$ for dry air leading to

$$\rho_{fluid} = \rho_{Air}(T) = \frac{p}{R_s \cdot T} \tag{3}$$

Where $p$ is the absolute pressure ($p = p_{static}$) and $R_s = 287 \, \frac{J}{kg \, K}$ is the specific gas constant.

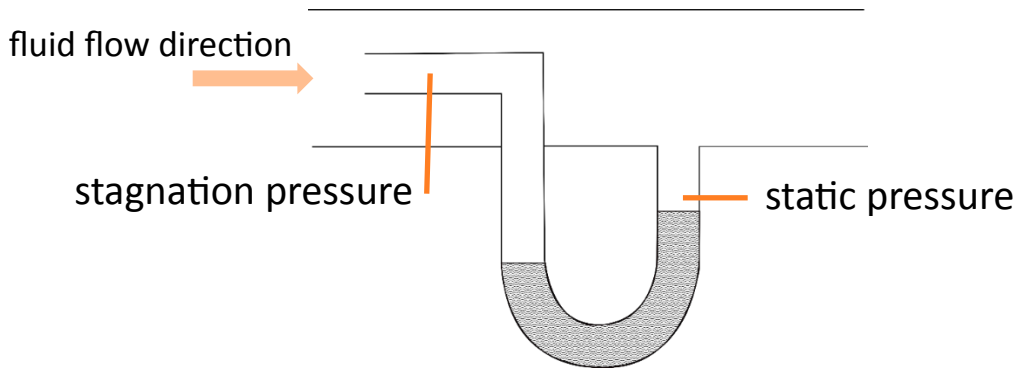

Figure 1: The Pitot tube in a cylinder with a moving fluid. The left opening of the Pitot tube is aligned against the flow direction and exposed to the stagnation pressure. The other end of the Pitot tube perpendicular to the fluid flow direction is exposed to the static pressure only, resulting in a height difference of the two fluid columns in the manometer.

## Experimental setup with internal pressure sensor

In the first experiment, we determine the speed of wind produced by a wind machine. Therefore, the static and stagnation pressure have to be measured. The static pressure is measured with the internal pressure sensor of the smartphone while the wind machine is switched off. Afterwards the wind machine is turned on and the same setup measures the stagnation pressure. Unlike the original Pitot tube, the pressure is measured with the smartphone's capacitive pressure sensor instead of the liquid fluid. The measurement process is illustrated in Figure 2, while a photo of the experimental setup can be seen in Figure 3.

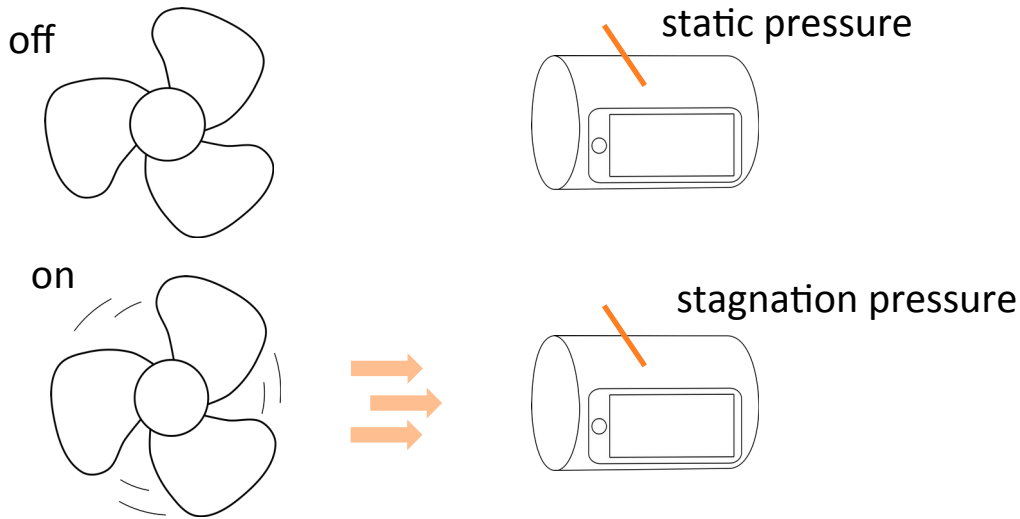

Figure 2: To determine the wind velocity, the static and stagnation pressure are measured with a smartphone, while the ventilator is switched off or after the ventilator is turned on, respectively.

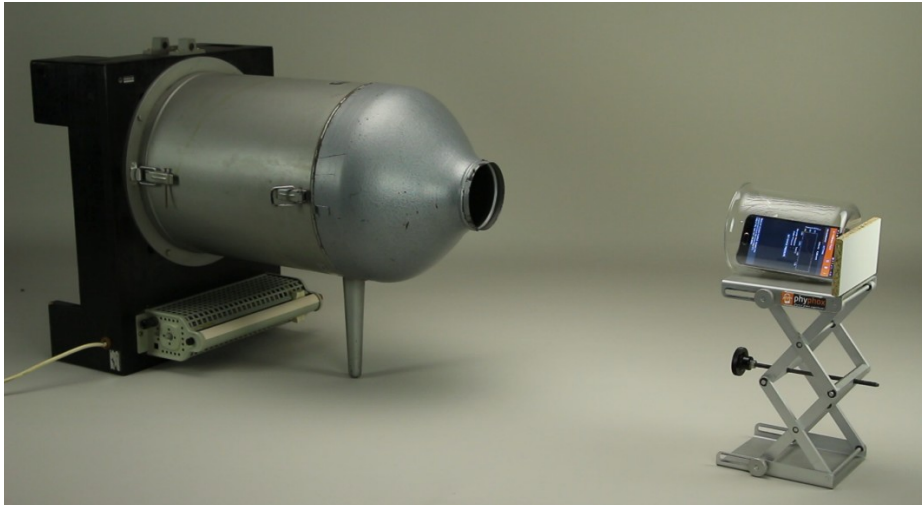

*Figure 3:* A picture of the experimental setup for measuring wind speed.

As described in Ref. 5, phyphox allows a customized in-app analysis of the measured sensor data and a user-designed display of its results. This can be created by any user and has also been used for the shown experiment. The resulting phyphox experiments are not available in phyphox by default, but can easily be shared with other users for example via a QR code to be scanned in phyphox (see Fig. 4).

Figure 4 shows the two screens "Measurement" and "Settings" in the designed phyphox experiment. At the bottom of the screen "Measurement", the current value of the measured pressure is shown. As long as the wind machine is switched off, this is the static pressure. By using the button "SET STATIC PRESSURE", the current numerical pressure value is set to the static pressure, which will be used for calculation according to equation (2). The air density, which is also required in equation (2), is calculated from the room temperature according to formula (3). The room temperature can be set in the settings tab, as shown in the right screenshot of Fig. 4.

If the wind machine is switched on, the current pressure value shows the stagnation pressure. It is given in both the lower graph of the pressure sensor's raw data and its current numerical value. These data can be converted to the air velocity by using equation (2) (see the top of the screen "Measurement" on the left side of Figure 4). Providing an appropriate choice of the experimental conditions, the upper graph shows meaningful velocity values. Exemplary measurement results are given in Figure 4. In this case, the wind machine was able to produce an air velocity of up to 10 m/s.

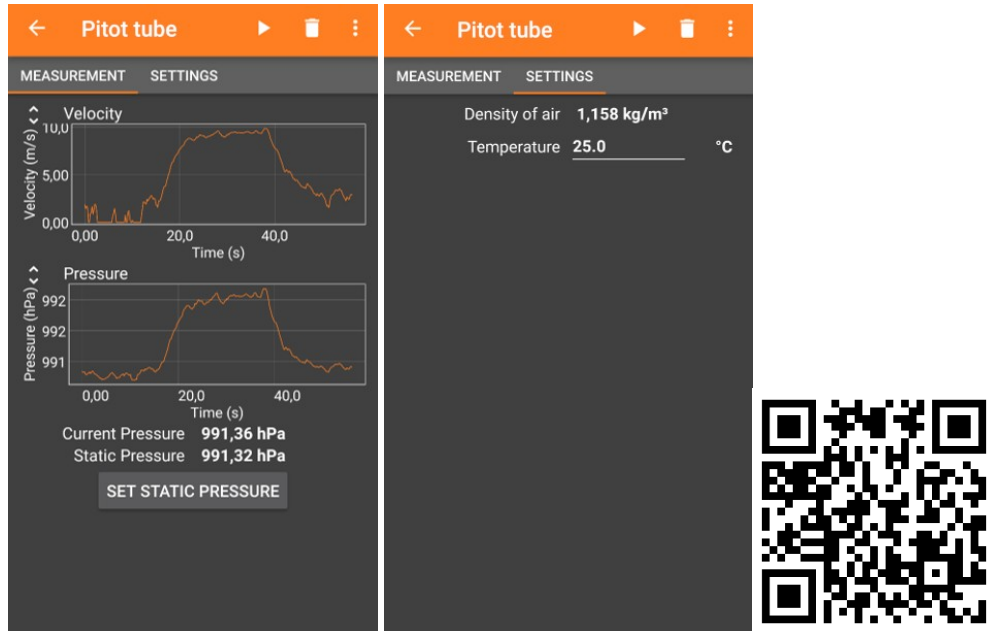

*Figure 4: Screenshots of a taken measurement. The calculated velocity is shown on top of the left screenshot. The graph below shows the measured raw data for the stagnation pressure. The values of the current pressure and the static pressure are shown below. To set the static pressure before the wind machine is turned on, the button "SET STATIC PRESSURE" is used. The room temperature is required for the calculation of the air density. It can be set in the settings tab, as shown in the middle screenshot. The experiment can be added to phyphox with the QR code on the right.*

## Experimental setup with external pressure sensors

Since version 1.1.0 the app phyphox can not only readout internal smartphone sensors but also external sensors via Bluetooth Low Energy (BLE). Thus, the described experiment can also be performed with a smartphone without an internal barometer by placing an external pressure sensor in the experimental setting in Figure 2 in place of the smartphone. As one can see in the sensor database *phyphoxDB*[13], roughly 48% of the *phyphoxDB* contributors have a pressure sensor available. Since the database is filled on a voluntary basis by phyphox users with their personally available sensors, the database is not necessarily a representative sample for students. However, as it clearly indicates that pressure sensors cannot be presumed to be available, external sensors support equal opportunities in performing student experiments. The use of external sensors can also reduce possible risks damaging smartphones in experiments. In addition, it facilitates the support of experimenting students if sensors with identical specifications are used in contrast to a possible broad variety of sensors in students' smartphones in the widespread BYOD approach (BYOD – bring your own device).

In our case of a Pitot tube experiment, using external sensors offers the additional possibility to improve the accuracy of the experiment by using two pressure sensors to measure the static and stagnation pressure simultaneously. This becomes important in all experimental settings where a time-dependent variation of the static pressure cannot be neglected. An example is using a Pitot tube to determine the velocity of a vehicle. Since streets are mostly not perfectly even, the static pressure is not constant anymore due to small variations in altitude. Hence, it is mandatory to measure stagnation and static pressure simultaneously.

Here, the pressure sensor of a Texas Instruments SensorTag CC2650 has been used as external sensor measuring the stagnation pressure. At the same time, the static pressure is measured with the internal pressure sensor of the smartphone. The phyphox experiment has been adapted to the modified experimental setup. The customized experiment can be added to phyphox by scanning the QR code in Figure 5.

The sensor module has been placed in a wide-necked bottle and held out of the window of a car while the car was in motion. The determined driving speed via the Pitot tube can be verified by comparing it to the velocity measured by a global navigation satellite system (GNSS), i.e. the GPS position of the smartphone. An exemplary measurement can be seen in Figure 5. The measured driving velocity from the Pitot tube in the top graph is noisier than the GPS generated data, but considering the simplicity of the experimental setup it fits very well to the measured GPS data. In this way, it validates the Bernoulli equation and demonstrates its applicability for velocity measurements in racing cars and aircrafts.

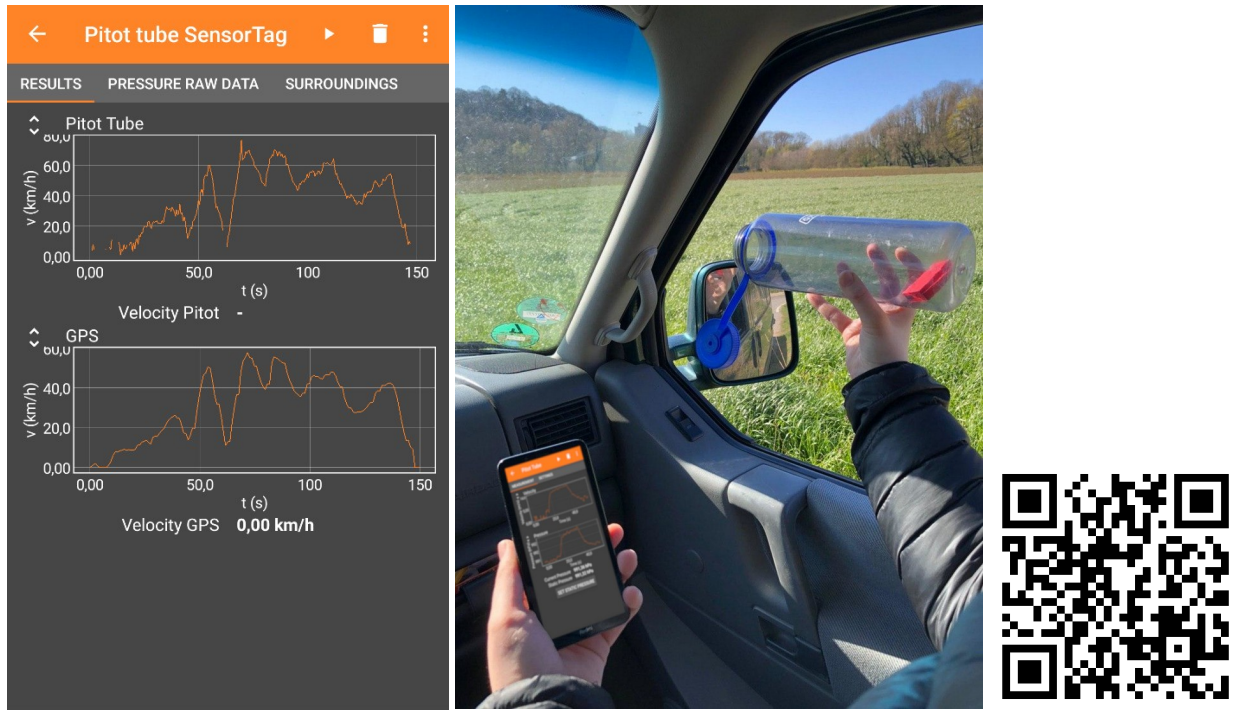

*Figure 5: Screenshot for the measurement of car velocity. The top graph shows the results of the measurement with a Pitot tube. The bottom graph presents the measured velocity by GPS. The results by the Pitot tube are noisier but fit overall very well with the GPS data. The customized experiment can be added to phyphox by scanning the QR code in the figure.*

## Conclusion

An experiment has been presented to measure the wind speed produced by a wind machine with a smartphone's barometer. By integrating external sensors, this experiment can be simply extended to measure the headwind while driving a car which can be used to determine a car's velocity. The velocity measured by the Pitot tube fits very well to the velocity determined from GPS data.

The experiments show both the possibility to design customized experiments by using the phyphox web-based editor and the extended experimental options by integrating external sensors into smartphone-based experiments. This opens a wide range of new experiments with smartphone-based data acquisition and analysis, which pursue a variety of educational goals not only for physics teaching, but for teaching science in general.

The following article has been accepted by "The Physics Teacher". After it is published, it will be found at https://aapt.scitation.org/journal/pte

# Acknowledgement

The Federal Ministry of Education and Research supports the project "Lehrerbildung Aachen" (LeBiAC2, FKZ 01JA1813) of the RWTH Aachen University in the context of their funding program "Qualitätsoffensive Lehrerbildung" (Phase 2). As part of LeBiAC2 we develop and evaluate the use of Bluetooth-based data acquisition by students during their education to become teachers.

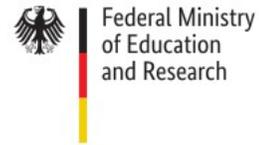